# Residue Classes of the PPT Sequence

Subhash Kak

**Abstract.** Primitive Pythagorean triples (PPT) may be put into different equivalence classes using residues with respect to primes. We show that the probability that the smaller odd number associated with the PPT triple is divisible by prime p is 2/(p+1). We have determined the autocorrelation function of the Baudhāyana sequences obtained from the residue classes and we show these sequences have excellent randomness properties. We provide analytical explanation for the peak and the average off-peak values for the autocorrelation function. These sequences can be used specifically in a variety of key generation and distribution problems and, more generally, as pseudorandom sequences.

**Introduction**
The theory of primitive Pythagorean triples (PPTs) [1],[2],[3] is related to other number theory areas such as Gaussian numbers, modular forms and spinors [4]. A Pythagorean triple (a, b, c) is a set of integers that are the sides of a right triangle and thus $a^2 + b^2 = c^2$. In this triple *a* is the smaller of the two odd numbers in the triple.

A primitive Pythagorean triple (PPT) consists of numbers that are relatively prime. To generate PPTs, one may start with different odd integers *s* and *t* that have no common factors and compute:

$$a = st; b = \frac{s^2 - t^2}{2}; c = \frac{s^2 + t^2}{2}$$

In earlier papers [2],[3] we considered properties of PPTs that had been put into 6 classes based on divisibility by 3, 4, and 5. Once the sequence of the PPTs has been listed in terms of the classes, one can determine the separation between elements of the same class and thus obtain a numerical sequence. We have named these sequences Baudhāyana sequences [3] after the ancient mathematician who described Pythagorean triples before Pythagoras [5],[6]. (For those not familiar with the historical context, please see [7]-[10].)

Here we consider properties of residue classes of the PPT sequence. We show that the probability that a PPT is divisible by prime p is 2/(p+1). We provide an analytical explanation for the general characteristics of the autocorrelation function and show that just like the sequences in 6 classes that were considered in [2], these sequences have excellent randomness properties.

## Residue Classes

We first prove a result on the number *a* (of the triple *a, b,* c) of PPTs for it to have zero residue with respect to mod *p*. This result is useful in determining the characteristics of the autocorrelation function.

**Theorem.** *The probability that a (of the triple a, b, c) of the PPTs is exactly divisible by prime p is* $\frac{2}{p+1}$.

*Proof.* PPTs are generated by the array shown below where the entries are odd.

|   |    | \multicolumn{6}{c}{s} |   |   |   |
|---|----|-------|-------|-------|-------|--------|--------|---|---|---|
|   |    | 3     | 5     | 7     | 9     | 11     | 13     | . | . | . |
|   | 1  | (1,3) | (1,5) | (1,7) | (1,9) | (1,11) | (1,13) | . | . | . |
|   | 3  |       | (3,5) | (3,7) |       | (3,11) | (3,13) | . | . | . |
|   | 5  |       |       | (5,7) | (5,9) | (5,11) | (5,13) | . | . | . |
| t | 7  |       |       |       | (7,9) | (7,11) | (7,13) | . | . | . |
|   | 9  |       |       |       |       | (9,11) | (9,13) | . | . | . |
|   | 11 |       |       |       |       |        | (11,13)| . | . | . |
|   | .  |       |       |       |       |        |        | . | . | . |
|   | .  |       |       |       |       |        |        | . | . | . |

**Table 1.** Array of PPTs ordered by *s* and *t* numbers [note that (3,9) is not a PPT]

For prime *p*, the number of a divisible entry along the horizontal and vertical directions will occur every *p* times. Of these $p^2$ items, exactly one is divisible by the corresponding *s* and *t* elements that are multiples of *p* and thus it does not qualify to be PPT; in other words we need to consider only $p^2 – 1$ items. There are *p - 1* items in one row and *p - 1* items in a column that are divisible by *p* for a total of *2( p – 1)* items. Therefore, in each such repeating two-dimensional pattern, the number of elements that are divisible by *p* divided by total elements that qualify as PPT is:

$$\frac{2(p-1)}{p^2-1} = \frac{2}{p+1}. \blacksquare$$

The value will be closer to the above as the number of points in the PPT sequences increases for then the effect of the items close to the diagonal will be reduced. A list of the first 100,000 PPTs that are exactly divisible by prime numbers in the range 3 through 29 is given in Table 2.

|  | Number of PPTs divisible by primes 3 through 29 | | | | | | | | |
|---|---|---|---|---|---|---|---|---|---|
| Divisor (*right*) Length (*below*) | 3 | 5 | 7 | 11 | 13 | 17 | 19 | 23 | 29 |
| 1000 | 514 | 341 | 255 | 180 | 153 | 116 | 116 | 80 | 80 |
| 2000 | 1035 | 672 | 505 | 384 | 324 | 227 | 227 | 171 | 117 |
| 3000 | 1537 | 1004 | 751 | 512 | 472 | 329 | 329 | 277 | 208 |
| 4000 | 2017 | 1349 | 992 | 674 | 595 | 462 | 381 | 320 | 240 |
| 5000 | 2772 | 1832 | 1405 | 947 | 797 | 640 | 544 | 448 | 376 |
| 10000 | 4997 | 3321 | 2481 | 1626 | 1431 | 1091 | 951 | 875 | 609 |
| 15000 | 7901 | 5192 | 3921 | 2619 | 2234 | 1818 | 1550 | 1372 | 1106 |
| 20000 | 10264 | 6857 | 5113 | 3327 | 2885 | 2255 | 2094 | 1773 | 1382 |
| 25000 | 12628 | 8424 | 6385 | 4248 | 3585 | 2883 | 2519 | 2110 | 1760 |
| 30000 | 15378 | 10241 | 7608 | 5130 | 4357 | 3424 | 2948 | 2581 | 1941 |
| 40000 | 21375 | 14267 | 10658 | 7247 | 6153 | 4745 | 4264 | 3513 | 2772 |
| 50000 | 26271 | 17503 | 13034 | 8783 | 7584 | 5792 | 5282 | 4544 | 3401 |
| 60000 | 30809 | 20554 | 15480 | 10110 | 8804 | 6828 | 6303 | 5157 | 3986 |
| 70000 | 35757 | 23842 | 17881 | 11836 | 10096 | 8072 | 7186 | 5836 | 4566 |
| 80000 | 40102 | 26732 | 20210 | 12393 | 11428 | 8848 | 7853 | 6598 | 5265 |
| 90000 | 45664 | 30441 | 22860 | 15217 | 13313 | 10199 | 9193 | 7753 | 6061 |
| 100000 | 50745 | 33845 | 25209 | 16769 | 14296 | 11046 | 9983 | 8625 | 6673 |

**Table 2.** Number of elements exactly divisible by primes in the top row (from 3 to 29)

For these primes, due to the boundary effect, the proportion would depart from the expected value of $\frac{2}{p+1}$ as the size of the prime increases. This is borne out by the entries of Table 2.

The problem in the divisibility of PPTs by composite numbers is slightly complicated as several other numbers in the group may share a divisor with the PPT. This interesting case will be taken up elsewhere.

### Baudhāyana sequences for residue classes

We now consider Baudhāyana sequences for residues with respect to primes 3, 5, and 7.

*Example 1.* First consider residues of *a* with respect to 3 in the increasing order of *c*.

    If  a mod 3 = 0; Class A,
    If  a mod 3 = 1; Class B,
    If  a mod 3 = 2; Class C

*The sequence obtained when indexed by increasing c:*

    ACAACAABAACACABAAACCACAAACAAABACACAABAAAACCAAACAAAABCAACCAC
    ABAACCAAAABAAAACABCACAAAAAAACCAAACCCABAAACAAACBAAACABCAACAC
    ABBAAAACACAAAAA……

The numerical sequence corresponding to As is: 2 1 2 1 2 1 2 2 2 1 1 … We can now compute the non-normalized autocorrelation function:

$$C(k) = \frac{1}{n+1} \sum_{i=0}^{n} a(i)a(i+k)$$

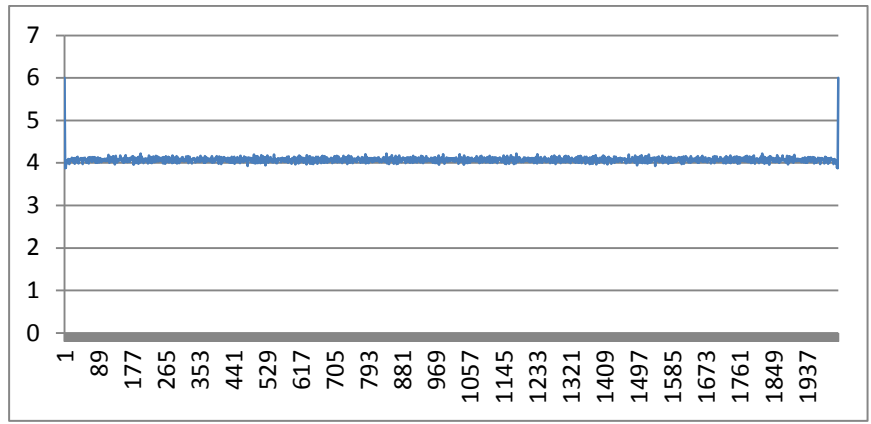

**Figure 1.** Autocorrelation of Class A for residue with respect to 3

For a sequence of length *n*, the number of A's in the sequence is *n/2*. Therefore, the average distance between two As is 2. The autocorrelation function for non-zero values of the argument to be approximately 4:

$$C(k) = E(a(i))E(a(i+k)) \approx 2 \times 2 \approx 4 \quad \text{(when } i \neq k\text{)}$$

The value of C(0) may be estimated by considering the separation between As as a random variable. It has a probability of ½ for 2 and for other values we may take the probability to go down as a power series of one-fourth, one-eighth, one-sixteenth, and so on.
C(0) = 1/2 × 4 + 1/4 × 1 + 1/8 × 9 + 1/16 × 16 + ... = 6.7

The actual value is somewhat less than this (Figure 1), which means that the assumption about the probability distribution made by us is not the best one.

If we consider the number of Bs and Cs in the sequence, the average separation between two Bs and Cs is 4. This makes the function for non-zero values of the argument to be approximately 16 as we find in Figure 2.

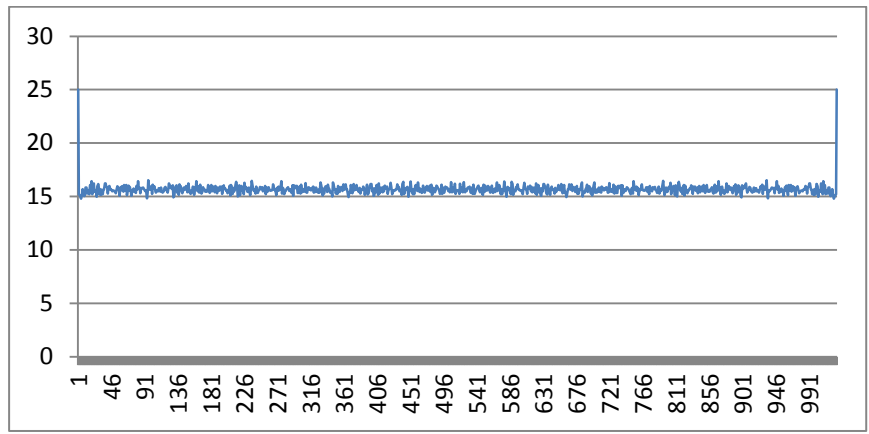

**Figure 2.** Autocorrelation of Class B for residue with respect to 3

*Example 2.* Consider residues of *a* with respect to 5 in the increasing order of *c*

    If  a mod 5 = 0; Class A,
    If  a mod 5 = 1; Class B,
    If  a mod 5 = 2; Class C,
    If  a mod 5 = 3; Class D,
    If  a mod 5 = 4; Class E

*The sequence obtained when indexed by increasing c:*

DAACBAEABDDADCEAEBACACDBAEAECDAACDBBBAEACDEABBACDAADDAEACC
AAAEAEBECCADCBABBEDDAACCEEAABDDADDBAABAACEBEAEAACCBBAABEDCA
CDAEBADDAAAEEEBDCAADCAEBECDDDEABBBACC……..

For a sequence of length *n*, the number of A's in the sequence is *n/3*. Therefore, the average distance between two As is 3. The function for non-zero values of the argument will be approximately 9 (see Figure 3). The function for non-zero values of the Bs, Cs, Ds and Es will be approximately 36.

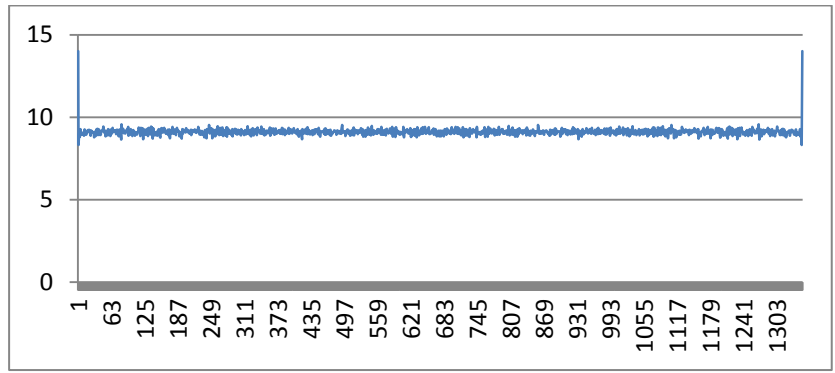

**Figure 3.** Autocorrelation of Class A for residue with respect to 5

*Example 3.* Consider residues of *a* with respect to 7 in the increasing order of *c*

If a mod 7 = 0; Class A,
If a mod 7 = 1; Class B,
If a mod 7 = 2; Class C,
If a mod 7 = 3; Class D,
If a mod 7 = 4; Class E,
If a mod 7 = 5; Class F,
If a mod 7 = 6; Class G

*The sequence obtained when indexed by increasing c:*
DFBAAACDEFAGGAECBABFADDCBAEFBGEGFAADEAGDCCGDAAFEAEFBBAFBGAC
CCADAGBDDFADBFCDECAGBGAEGBAFFCEEDABDFAACCFAABADCCBGDAAAEEEB
GAEFGAGFACDCGGEECGDAAFDDAFBBDFGEAG………..

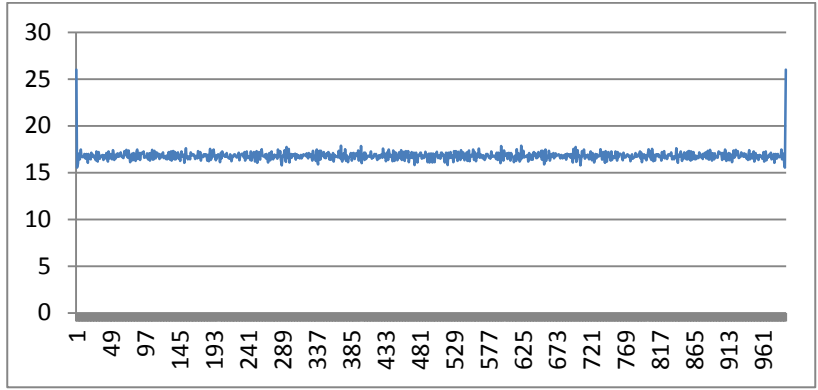

**Figure 4.** Autocorrelation of Class A for residue with respect to 7

For a sequence of length *n*, the number of A's in the sequence is n/4. Therefore, the average distance between two As is 4. The function for non-zero values of the argument will be approximately 16 (Figure 4). The function for non-zero values of the Bs, Cs, Ds, Es, Fs and Gs will be approximately 64.

## Conclusions

The paper shows that when we divide the PPTs into residue classes they exhibit excellent randomness properties for all the classes. We show that the probability that *a* (of the PPT triple *a,b,c*) is divisible by prime *p* is *2/(p+1)*. We have determined the autocorrelation function of the Baudhāyana sequences obtained from the residue classes and we show these sequences have excellent randomness properties. We provide analytical explanation for the peak and the average off-peak values for the autocorrelation function. Since PPTs can be seen most naturally as two-dimensional arrays (with holes when the gcd is not 1), the work reported here can be a different starting point for array computation [11]-[14].

The research in this paper can also be seen from the interesting perspective of considering primality with respect to some specified non-geometrical property of an array of odd numbers. We stress "non-geometrical" since primality can be triviality defined in terms of prime number of regular partitions of an array. In non-geometrical partition of the kind considered here for the *s* and *t* values of Table 1, the equivalence classes with respect to different primes are not defined in an obvious geometric way.

**Acknowledgement**. I thank Monisha Prabhu for helping with the computations associated with this research.